\documentclass[11pt]{article}
\usepackage{amssymb,amsmath,amsfonts}
\usepackage{graphicx}
\usepackage{graphics}
\usepackage{eepic,epsfig}

\begin{document}

\title{Surface Casimir densities on branes orthogonal to the boundary \\ of
anti-de Sitter spacetime}
\author{A. A. Saharian\thanks{%
E-mail: saharian@ysu.am } \\
%EndAName
\textit{Institute of Physics, Yerevan State University, }\\
\textit{1 Alex Manogian Street, 0025 Yerevan, Armenia }}
\maketitle

\begin{abstract}
We investigate the vacuum expectation value of the surface energy-momentum
tensor (SEMT) for a scalar field with general curvature coupling in the
geometry of two branes orthogonal to the boundary of anti-de Sitter (AdS)
spacetime. For Robin boundary conditions on the branes, the SEMT is
decomposed into the contributions corresponding to the self-energies of the
branes and the parts induced by the presence of the second brane. The
renormalization is required for the first parts only and for the
corresponding regularization the generalized zeta function method is
employed. The induced SEMT is finite and is free from renormalization
umbiguities. For an observer living on the brane, the corresponding equation of state is of the cosmological constant type. Depending on the boundary conditions and on the separation between the branes, the surface energy densities can be either positive or negative. The energy density induced on the brane vanishes in special cases of Dirichlet and Neumann boundary conditions on that brane. The effect of gravity on the induced SEMT is essential at separations between the branes of the order or larger than the curvature radius for AdS spacetime. In the large separation limit the decay of the SEMT, as a function of the proper separation, follows a power law for both massless and massive fields. For parallel plates in Minkowski bulk and for massive fields the fall-off of the corresponding expectation value is exponential.
\end{abstract}

\bigskip


\bigskip

\section{Introduction}

Among the interesting direction in the developments of the Casimir effect
theory (for general introduction and applications see, e.g., \cite{Most97}-%
\cite{Casi11}) is the study of dependence of expectation values of physical
characteristics for quantum fields on the bulk and boundary geometries, as
well as on the spatial topology. The interest is motivated by applications
in gravitational physics, in cosmology and in condensed matter physics.
Exact analytic expressions for physical characteristics are obtained in
geometries with a sufficient degree of symmetry. In particular, the
respective background geometries include maximally symmetric spacetimes
sourced by positive and negative cosmological constants. These geometries,
referred as de Sitter (dS) and anti-de Sitter (AdS) spacetimes,
respectively, are among the most popular bulks in quantum field theory on
curved backgrounds.

The goal of this paper is to investigate the surface Casimir densities on
two parallel branes for a scalar field in AdS spacetime. Quantum field
theoretical effects on fixed AdS background have been extensively studied in
the literature. The importance of those investigations is motivated by
several reasons. The AdS spacetime is a non-globally hyperbolic manifold
with a timelike boundary at spatial infinity and the early interest to the
formulation of quantum field theory in that geometry was related to
principal questions of quantization \cite{Avis78,Brei82,Mezi85} (see also
the references in \cite{Saha05}). The necessity to control the information
through the spatial infinity requires the imposition of boundary conditions
on quantum fields (for a discussion of possible boundary conditions on the
AdS boundary see, e.g., \cite{Ishi04,Morl22}). The different boundary
conditions correspond to physically different field theories. The AdS
boundary at spatial infinity plays a central role in models of AdS/Conformal
Field Theory (AdS/CFT) correspondence \cite{Ahar00}-\cite{Ammo15}. The
latter establishes duality between conformal field theory living on the
boundary of AdS spacetime and supergravity or string theory on AdS bulk.
This holographic correspondence between two different theories provides an
efficient computational framework for non-perturbative effects, mapping them
to the perturbative region of the dual theory. Within this approach
interesting results have been obtained in high energy physics, in quantum
chromodynamics and in condensed matter physics \cite{Papa11,Pire14,Zaan15}.
The braneworld models \cite{Maar10} with large extra dimensions, both
phenomenological and string theory motivated, present another interesting
setup where the properties of AdS spacetime play a crucial role. They
provide a geometrical solution to the hierarchy problem between the
electroweak and gravitational energy scales and serve as an interesting
framework to discuss the problems in high energy physics, gravitation and
cosmology.

The braneworld models contain two types of fields: fields propagating in the
bulk and fields localized on the branes. In simplified models, the
interaction between branes and bulk fields is reduced to boundary conditions
on the branes. Those conditions modify the spectrum of vacuum fluctuations
of bulk quantum fields and give rise to the Casimir type contributions in
the expectation values of physical observables, such as the ground state
energy and the vacuum forces acting on the branes. The Casimir energy and
forces in the geometry of branes parallel to the AdS boundary have been
widely studied in the literature (see \cite{Fabi00}-\cite{Moss03} for early
investigations and \cite{Saha20Rev} for a more complete list of references).
The Casimir forces can be used as a possible mechanism for stabilization of
interbrane distance that is required to escape the variations of physical
constants in the effective theory on the branes. The vacuum fluctuations of
bulk field may also provide a mechanism for generation of cosmological
constant on branes. More detailed information on the properties of the
vacuum state is contained in the expectation values of bilinear combinations
of fields, such as the field squared and the energy-momentum tensor. In
braneworld models on AdS bulk those expectation values are considered in
\cite{Saha03}, \cite{Knap04}-\cite{Saha20} for scalar, fermionic and
electromagnetic fields. For charged fields, another important local
characteristic of the vacuum state is the expectation value of the current
density. The combined effects of branes and spatial topology on the vacuum
currents for scalar and fermionic fields in locally AdS spacetime, with a
part of spatial dimensions compactified on a torus, have been studied in
\cite{Beze15a}-\cite{Bell20}.

In the references cited above the branes are parallel to the AdS boundary
(Randall-Sundrum-type models \cite{Rand99a,Rand99b}). In a number of recent
developments in conformal field theories additional boundaries are present
(see, e.g., \cite{Cuom21} and references therein). In the context of AdS/CFT
correspondence, the respective dual theory on the AdS bulk contains
boundaries intersecting the AdS boundary (AdS/BCFT correspondence) \cite%
{Taka11,Fuji11}. Another interesting problem on AdS bulk with surfaces
crossing its boundary is related to the evaluation of the entanglement
entropy of a quantum system in conformal field theory with a boundary. In
accordance of the procedure suggested in \cite{Ryu06,Ryu06b}, the
entanglement entropy in a bounded region from the CFT side on the AdS
boundary is expressed in terms of the area of the minimal surface in the AdS
bulk that asymptotes the boundary of CFT (see also \cite{Nish09,Chen22} for
reviews). Motivated by those developments, in \cite{Beze15,Bell22} we have
studied the influence of branes, orthogonally intersecting the AdS boundary,
on the local properties of the scalar vacuum in general number of spatial
dimensions. As local characteristics of the vacuum state, the expectation
values of the field squared and of the energy-momentum tensor have been
considered. By using the respective vacuum stresses, the Casimir forces
acting on the branes were investigated as well. It has been shown that, in
addition to the component perpendicular to the brane, those forces have a
nonzero parallel component (shear force). In quantum field theory with
boundaries the expectation values of physical quantities may contain
contributions localized on the boundary. The expression for the surface
energy-momentum tensor of a scalar field with general curvature coupling
parameter and for general bulk and boundary geometries has been derived in
\cite{Saha04} by using the standard variational procedure. The corresponding
vacuum expectation value in the problem with branes parallel to the AdS
boundary is investigated in \cite{Saha04cc,Saha06cc}. The present paper
considers the vacuum expectation value of the surface energy-momentum tensor (SEMT)
for a scalar field in the problem with two parallel branes orthogonal to the
AdS boundary.

The organization of the paper is as follows. In the next section we describe
the geometry of the problem and present the expression for the surface
energy-momentum tensor. The corresponding vacuum expectation value (VEV) is
investigated in Section \ref{sec:VEV} by using the two-point function from
\cite{Bell22}. The surface energy density is decomposed into contributions
corresponding to the self-energy of the brane when the second brane is
absent and the part induced by the second brane. The renormalization is
required only for the first contribution. In the limit of infinite curvature
radius we recover the result for parallel plates in Minkowski bulk. Another
special case with conformal relation to the Casimir problem in Minkowski
spacetime corresponds to a conformally coupled massless field. The behavior
of the SEMT in asymptotic regions of the
parameters is discussed in Section \ref{sec:Asymp}. The numerical analysis
for the induced surface energy density is presented as well. The main
results of the paper are summarized in Section \ref{sec:Conc}. The
regularization of the self-energy contribution, by using the generalized
zeta function approach, is considered in Appendix \ref{sec:App}. The finite
part is separated on the basis of principal part prescription.

\section{Geometry of the problem}

\label{sec:Geom}

AdS spacetime is the maximally symmetric solution of the Einstein equations
with a negative cosmological constant $\Lambda $ as the only source of the
gravitational field. In Poincar\'{e} coordinates $(t,x^{1},\mathbf{x},z)$,
with $\mathbf{x}=(x^{2},\ldots ,x^{D-1})$ and $D$ being the number of
spatial dimensions, the respective metric tensor $g_{ik}$ is given by%
\begin{equation}
ds^{2}=g_{ik}dx^{i}dx^{k}=\left( \frac{\alpha }{z}\right) ^{2}\left[
dt^{2}-\left( dx^{1}\right) ^{2}-d\mathbf{x}^{2}-dz^{2}\right] .  \label{dsz}
\end{equation}%
Here, the parameter $\alpha =\sqrt{D(1-D)/(2\Lambda )}$ determines the
curvature radius of the background spacetime, $-\infty <x^{i}<+\infty $ for $%
i=0,1,2,\ldots ,D-1$, and $0\leq z<\infty $. The $D$-dimensional
hypersurfaces $z=0$ and $z=\infty $ present the AdS boundary and horizon,
respectively. The proper distance along the $z$-direction is measured by the
coordinate $y=\alpha \ln (z/\alpha )$, $-\infty <y<+\infty $. In the
coordinate system $(t,x^{1},\mathbf{x},y)$ one has $g_{DD}^{\prime }=1$ and $%
g_{ik}^{\prime }=g_{ik}=e^{-2y/\alpha }\eta _{ik}$, $i,k=0,1,\ldots ,D-1$,
with $\eta _{ik}$ being the metric tensor for Minkowski spacetime.

\begin{figure}[tbph]
\begin{center}
\epsfig{figure=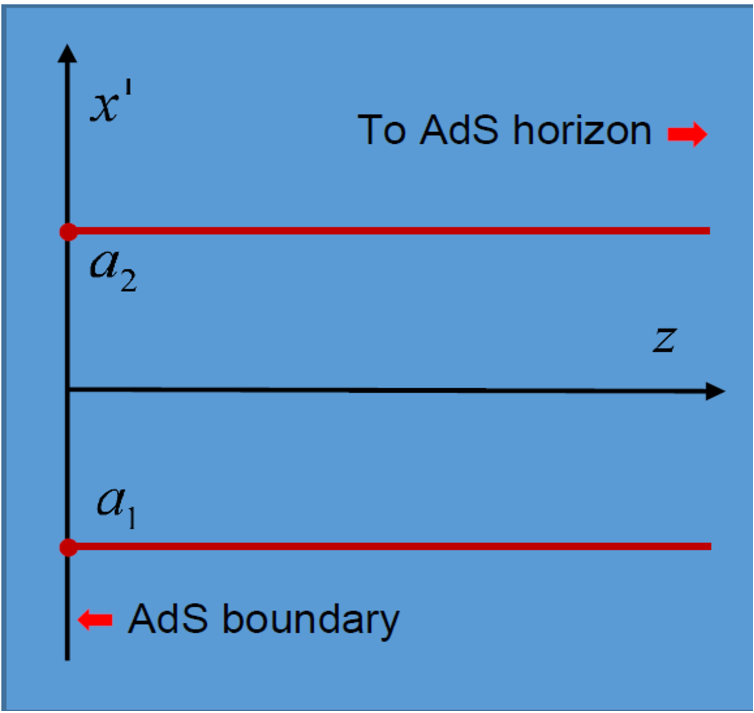,width=5.5cm,height=5.5cm}
\end{center}
\caption{The geometry of two branes orthogonal to the AdS boundary. }
\label{fig1}
\end{figure}

We aim to investigate the surface Casimir densities induced by quantum
fluctuations of a scalar field $\varphi (x)$ on codimension one parallel
branes located at $x^{1}=a_{1}$ and $x^{1}=a_{2}$, $a_{1}<a_{2}$ (see Figure %
\ref{fig1} for the geometry of the problem). It will be assumed that the
field is prepared in the Poincar\'{e} vacuum state. For a scalar field with
curvature coupling parameter $\xi $ the corresponding field equation reads
\begin{equation}
\left( \square +\xi R+m^{2}\right) \varphi (x)=0,  \label{feq}
\end{equation}%
where $\square =g^{ik}\nabla _{i}\nabla _{k}$ is the covariant d'Alembertian
and $R=2\Lambda (D+1)/(D-1)$ is the Ricci scalar for AdS spacetime. On the
branes, the field operator is constrained by Robin boundary conditions%
\begin{equation}
(A_{j}+B_{j}n_{(j)}^{i}\nabla _{i})\varphi (x)=0,\;x^{1}=a_{j},  \label{Rbc}
\end{equation}%
where $n_{(j)}^{i}$ is the normal to the brane at $x^{1}=a_{j}$ pointing
into the region under consideration. The branes divide the background space
into three regions: $x^{1}\leq a_{1}$, $a_{1}\leq x^{1}\leq a_{2}$, and $%
x^{1}\geq a_{2}$. In the first and third regions one has $%
n_{(1)}^{i}=-\delta _{1}^{i}z/\alpha $ and $n_{(2)}^{i}=\delta
_{1}^{i}z/\alpha $, respectively. For the region $a_{1}\leq x^{1}\leq a_{2}$
the normal in (\ref{Rbc}) is expressed as $n_{(j)}^{i}=(-1)^{j-1}\delta
_{1}^{i}z/\alpha $. In the discussion below we consider the region between
the branes. The VEVs for the regions $x^{1}\leq a_{1}$ and $x^{1}\geq a_{2}$
are obtained in the limits $a_{2}\rightarrow \infty $ and $a_{1}\rightarrow
-\infty $. For the sets of the coefficients $(A_{j},B_{j})=(A_{j},0)$ and $%
(A_{j},B_{j})=(0,B_{j})$ the constraints (\ref{Rbc}) are reduced to
Dirichlet and Neumann boundary conditions, respectively. For Robin boundary
conditions, here the special case $B_{j}/A_{j}=\alpha \beta _{j}/z$ will be
assumed with $\beta _{j}$, $j=1,2$, being constants. For this choice, the
boundary conditions (\ref{Rbc}), written in terms of the coordinate $%
x_{(p)}^{1}=\alpha x^{1}/z$, take the form%
\begin{equation}
(1+\beta _{j}n_{(j)}^{1}\partial _{x_{(p)}^{1}})\varphi (x)=0,\;x^{1}=a_{j}.
\label{Rbc2}
\end{equation}%
The latter is the Robin boundary condition with constant coefficient $\beta
_{j}$. This coefficient characterizes the properties of the brane and can be
used to model the finite penetrations length of quantum fluctuations. Note
that the coordinate $x_{(p)}^{1}$ in (\ref{Rbc2}) measures the proper
distance from the brane for fixed $z$.

For the scalar field modes in the region between the branes the eigenvalues
of the quantum number $k^{1}$, corresponding to the momentum along the
direction $x^{1}$, are quantized by the boundary conditions (\ref{Rbc2}).
Those eigenvalues are roots of the transcendental equation (see \cite{Bell22}%
)%
\begin{equation}
\left( \beta _{1}+\beta _{2}\right) k^{1}a\cos \left( k^{1}a\right) +\left[
\beta _{1}\beta _{2}(k^{1})^{2}-1\right] \sin \left( k^{1}a\right) =0,
\label{Eigval}
\end{equation}%
where $a=a_{2}-a_{1}$. Depending on the values of the Robin coefficients
this equation, in addition to an infinite set of roots with real $k^{1}$,
may have purely imaginary roots $k^{1}=i\chi $ (for the corresponding
conditions see \cite{Rome02}). The energy of the scalar modes, with the
momentum $\mathbf{k}=(k^{2},\dots ,k^{D-1})$, $-\infty <k^{i}<+\infty $, $%
i=2,\ldots ,D-1$, in the subspace with coordinates $\mathbf{x}$, is
expressed as $E=\sqrt{(k^{1})^{2}+\mathbf{k}^{2}+\gamma ^{2}}$, where $0\leq
\gamma <\infty $ is the quantum number corresponding to the $z$-direction.
The dependence of the mode functions on the coordinate $z$ is expressed in
terms of the function $z^{D/2}J_{\nu }(\gamma z)$, with $J_{\nu }(u)$ being
the Bessel function and%
\begin{equation}
\nu =\sqrt{\frac{D^{2}}{4}-D(D+1)\xi +m^{2}\alpha ^{2}}.  \label{nuJ}
\end{equation}%
Note that, in contrast to the Minkowski bulk, the energy of the scalar modes
with given momentum does not depend on the mass of the field quanta. The
mass enters in the problem through the parameter $\nu \geq 0$. Now, we see
that in the presence of imaginary roots $k^{1}=i\chi $, for the scalar field
modes with $\mathbf{k}^{2}+\gamma ^{2}<\chi ^{2}$ the energy becomes
imaginary. This signals about the instability of the vacuum state under
consideration. In the discussion below we will assume the values of the
coefficients $\beta _{1}$ and $\beta _{2}$ for which there are no imaginary
roots of the eigenvalue equation (\ref{Eigval}). The corresponding
conditions read \cite{Rome02}
\begin{equation}
\beta _{1,2}\leq 0\cup \{\beta _{1}\beta _{2}\leq 0,\beta _{1}+\beta
_{2}>1/a\}.  \label{StabCond}
\end{equation}

For a general $(D+1)$-dimensional spacetime with a smooth boundary $\partial
M_{s}$, the SEMT $T_{ik}^{\mathrm{(s)}%
}(x)=\tau _{ik}\delta (x;\partial M_{s})$, localized on the boundary by the
one-sided delta function $\delta (x;\partial M_{s})$, is given by \cite%
{Saha04}%
\begin{equation}
\tau _{ik}=(1/2-2\xi )h_{ik}\varphi n^{l}\nabla _{l}\varphi +\xi
K_{ik}\varphi ^{2}.  \label{tauik}
\end{equation}%
Here, $h_{ik}=g_{ik}+n_{i}n_{k}$ is the induced metric on the boundary, with
$n_{i}$ being the inward-pointing unit normal vector for $\partial M_{s}$,
and $K_{ik}=h_{i}^{l}h_{k}^{m}\nabla _{l}n_{m}$ is the respective extrinsic
curvature tensor. The expression (\ref{tauik}) was obtained in \cite{Saha04}
by using the standard variational procedure for the action of a scalar field
with general curvature coupling parameter and with an appropriate boundary
term localized on $\partial M_{s}$. Denoting the vacuum state by $|0\rangle $%
, the VEV of the SEMT is presented as
\begin{equation}
\langle 0|T_{ik}^{\mathrm{(s)}}|0\rangle =\delta (x;\partial M_{s})\langle
0|\tau _{ik}|0\rangle ,  \label{Semtvev}
\end{equation}%
where the VEV $\langle \tau _{ik}\rangle \equiv \langle 0|\tau
_{ik}|0\rangle $ is written in terms of the Hadamard function $%
G^{(1)}(x,x^{\prime })=\langle 0|\varphi (x)\varphi (x^{\prime })+\varphi
(x^{\prime })\varphi (x)|0\rangle $ by the formula%
\begin{equation}
\langle \tau _{ik}(x)\rangle =\frac{1}{2}\lim_{x^{\prime }\rightarrow x}%
\left[ (1/2-2\xi )h_{ik}n^{l}\nabla _{l}+\xi K_{ik}\right]
G^{(1)}(x,x^{\prime }).  \label{tauikvev}
\end{equation}%
The limit in the right-hand side contains two types of divergences. The
first one is present already in the case when the point $x$ does not belong
to the boundary. The corresponding divergent part is the same as that in the
problem where the branes are absent and it is removed by the subtraction
from the Hadamard function in (\ref{tauikvev}) the corresponding function in
the brane-free geometry. The SEMT is absent in the latter geometry and the
brane-free Hadamard function does not contribute to the VEV of the SEMT. The
second type of divergences originates from the surface divergences in
quantum field theory with boundaries and arise when the point $x$ belongs to
the boundary.

\section{VEV of the SEMT}

\label{sec:VEV}

\subsection{General expression}

In the problem under consideration and for the region $a_{1}\leq x^{1}\leq
a_{2}$ the inward-pointing normal is given by $n_{i}=n_{(j)i}=(-1)^{j}\delta
_{i}^{1}\alpha /z$ for the brane at $x^{1}=a_{j}$. The corresponding induced
metric reads $h_{ik}=g_{ik}$,$\;i,k\neq 1$, and $h_{11}=0$. Now, it can be
easily checked that the extrinsic curvature tensor for the branes vanishes, $%
K_{ik}=0$. Hence, the VEV of the SEMT is expressed as%
\begin{equation}
\langle \tau _{ik}(x)\rangle =\left( \frac{1}{4}-\xi \right)
h_{ik}n^{l}\lim_{x^{\prime }\rightarrow x}\nabla _{l}G^{(1)}(x,x^{\prime }).
\label{tauikvev2}
\end{equation}%
The expression for the Hadamard function in the region between the branes is
obtained from the corresponding expression for the Wightman function derived
in \cite{Bell22}. It is presented in the decomposed form%
\begin{eqnarray}
G^{(1)}(x,x^{\prime }) &=&G_{j}^{(1)}(x,x^{\prime })+\frac{2(zz^{\prime })^{%
\frac{D}{2}}}{(2\pi \alpha )^{D-1}}\int d\mathbf{k\,}e^{i\mathbf{k}\Delta
\mathbf{x}}\int_{0}^{\infty }d\gamma \,\gamma J_{\nu }(\gamma z)J_{\nu
}(\gamma z^{\prime })  \notag \\
&&\times \int_{w}^{\infty }d\lambda \frac{\cosh (\sqrt{\lambda ^{2}-w^{2}}%
\Delta t)}{\sqrt{\lambda ^{2}-w^{2}}}\frac{2\cosh \left[ {\lambda }\left(
x^{1}-x^{\prime 1}\right) \right] {+}\sum_{l=\pm 1}\left[ {e}^{{%
|x^{1}+x^{\prime 1}{-2{a_{j}}}|\lambda }}c_{j}(\lambda )\right] ^{l}}{%
c_{1}(\lambda )c_{2}(\lambda )e^{2a\lambda }-1},  \label{G1}
\end{eqnarray}%
where $\Delta \mathbf{x}=\mathbf{x}-\mathbf{x}^{\prime }$, $w=\sqrt{\gamma
^{2}+k^{2}}$, $k=|\mathbf{k}|$, and%
\begin{equation}
c_{j}(\lambda )=\frac{\beta _{j}\lambda -1}{\beta _{j}\lambda +1}.
\label{cj}
\end{equation}%
In (\ref{G1}),
\begin{eqnarray}
G_{j}^{(1)}(x,x^{\prime }) &=&G_{0}^{(1)}(x,x^{\prime })+\frac{(zz^{\prime
})^{D/2}}{(2\pi \alpha )^{D-1}}\int d\mathbf{k\,}e^{i\mathbf{k}\Delta
\mathbf{x}}\int_{0}^{\infty }d\gamma \,\gamma J_{\nu }(\gamma z)J_{\nu
}(\gamma z^{\prime })  \notag \\
&&\times \int_{0}^{\infty }d\lambda \,\frac{e^{-i\sqrt{\lambda ^{2}+w^{2}}%
\Delta t}}{\sqrt{\lambda ^{2}+w^{2}}}\sum_{l=\pm 1}\left[ {e}^{i{%
|x^{1}+x^{\prime 1}{-2{a_{j}}}|\lambda }}c_{j}(i\lambda )\right] ^{l},
\label{Gj}
\end{eqnarray}%
is the Hadamard function in the problem with a brane at $x^{1}=a_{j}$ when
the second brane is absent. Again, it is obtained from the respective
Wightman function given in \cite{Beze15,Bell22}. The first term in the
right-hand side, $G_{0}^{(1)}(x,x^{\prime })$, is the Hadamard function in
AdS spacetime without branes. The last term in (\ref{G1}) is interpreted as
the contribution to the Hadamard function in the region $a_{1}\leq x^{1}\leq
a_{2}$, induced by the brane at $x^{1}=a_{j^{\prime }}$ when we add it to
the problem with a single brane at $x^{1}=a_{j}$. Here and below, $j^{\prime
}=1$ for $j=2$ and $j^{\prime }=2$ for $j=1$.

Combining (\ref{tauikvev2}) and (\ref{G1}), the SEMT on the brane at $%
x^{1}=a_{j}$ is decomposed as%
\begin{equation}
\langle \tau _{ik}\rangle _{j}=\langle \tau _{ik}\rangle _{j}^{(0)}+\langle
\tau _{ik}\rangle _{j}^{\mathrm{ind}}.  \label{taudec}
\end{equation}%
Here, $\langle \tau _{ik}\rangle _{j}^{(0)}$ is the VEV of the SEMT when the
second brane is absent and $\langle \tau _{ik}\rangle _{j}^{\mathrm{ind}}$
is induced by the second brane at $x^{1}=a_{j^{\prime }}$. The VEV $\langle
\tau _{ik}\rangle _{j}^{(0)}$ is obtained from (\ref{tauikvev2}) with the
Hadamard function (\ref{Gj}). By taking into account that in the AdS
spacetime without branes the SEMT is absent, we get%
\begin{equation}
\langle \tau _{i}^{k}\rangle _{j}^{(0)}=(4\xi -1)\frac{\delta _{i}^{k}\beta
_{j}z^{D+1}}{(2\pi )^{D-1}\alpha ^{D}}\int d\mathbf{k\,}\int_{0}^{\infty
}d\gamma \,\gamma J_{\nu }^{2}(\gamma z)\int_{0}^{\infty }d\lambda \,\frac{1%
}{\sqrt{\lambda ^{2}+b^{2}}}\frac{\lambda ^{2}}{1+\lambda ^{2}\beta _{j}^{2}}%
.  \label{tau0}
\end{equation}%
The vacuum SEMT induced by the second brane comes from the last term in (\ref%
{G1}). It is presented in the form%
\begin{eqnarray}
\langle \tau _{i}^{k}\rangle _{j}^{\mathrm{ind}} &=&(4\xi -1)\frac{2\delta
_{i}^{k}\beta _{j}z^{D+1}}{(2\pi )^{D-1}\alpha ^{D}}\int d\mathbf{k\,}%
\int_{0}^{\infty }d\gamma \,\gamma J_{\nu }^{2}(\gamma z)\int_{b}^{\infty
}d\lambda \frac{\lambda ^{2}}{\sqrt{\lambda ^{2}-b^{2}}}  \notag \\
&&\times \frac{\beta _{j^{\prime }}\lambda +1}{\beta _{j}\lambda -1}\frac{1}{%
\left( \beta _{1}\lambda -1\right) \left( \beta _{2}\lambda -1\right)
e^{2a\lambda }-\left( \beta _{1}\lambda +1\right) \left( \beta _{2}\lambda
+1\right) }.  \label{tauind}
\end{eqnarray}%
The expression (\ref{tau0}) for the self-SEMT is divergent and needs a
regularization with a subsequent renormalization removing the divergences.
This type of surface divergences are well known in quantum field theory with
boundaries.

Note that for an observer living on the brane $x^{1}=a_{j}$ the $D$%
-dimensional line element is obtained from (\ref{dsz}) taking $dx^{1}=0$. It
describes $D$-dimensional AdS spacetime generated by a cosmological constant
$\Lambda ^{\prime }=(1-2/D)\Lambda $. From the point of vew of an observer
on the brane, the energy-momentum tensor $\langle \tau _{i}^{k}\rangle _{j}$
is a source of gravitation with the energy density $\varepsilon _{j}=\langle
\tau _{0}^{0}\rangle _{j}$ and isotropic effective pressure $p_{j}=-\langle
\tau _{2}^{2}\rangle _{j}=\cdots =-\langle \tau _{D}^{D}\rangle _{j}$. The
corresponding equation of state reads $p_{j}=-\varepsilon _{j}$ and, hence, $%
\langle \tau _{i}^{k}\rangle _{j}$ is a source of the cosmological constant
type. Of course, the latter property is a consequence of the symmetry in the
problem under consideration. In accordance with (\ref{taudec}), the surface
energy density is decomposed into the self-energy and the contribution
induced by the second brane:%
\begin{equation}
\varepsilon _{j}=\varepsilon _{j}^{(0)}+\varepsilon _{j}^{\mathrm{ind}},
\label{epsjdec}
\end{equation}%
where $\varepsilon _{j}^{\mathrm{ind}}=\langle \tau _{0}^{0}\rangle _{j}^{%
\mathrm{ind}}$.

The regularization of the divergent expression in the right-hand side of (%
\ref{tau0}), based on the generalized zeta function approach, is discussed
in appendix \ref{sec:App}. It is decomposed into pole and finite
contributions obtained from (\ref{Fdec}) in combination with (\ref{analytic}%
). In the principal part prescription the finite self-energy $\varepsilon
_{j}^{(0)}$ is identified with the finite part of the respective Laurent
expansion near the physical point $s=1$. In order to remove the divergent
part we note that the VEV $\langle \tau _{ik}\rangle _{j}$ is a part of a
theory which contains other contributions localized on the brane and the
divergences in $\langle \tau _{ik}\rangle _{j}$ are absorbed by
renormalizing the parameters in those contributions. The finite part of the
SEMT $\langle \tau _{ik}\rangle _{j}^{(0)}$ is given by (\ref{tau0fin}).
This part contains renormalization umbiguities which can be fixed by
imposing additional renormalization conditions. Here the situation is
completely parallel to that for the total Casimir energy discussed, for
example, in \cite{Bord09}. Similar to (\ref{taudec}), the Casimir energy for
a system composed of separate bodies is decomposed into the self energies
and the interaction energy. The renormalization is required only for the
self energies

Unlike to the self energy part $\varepsilon _{j}^{(0)}$, the surface energy
density $\varepsilon _{j}^{\mathrm{ind}}$ and the related SEMT $\langle \tau
_{i}^{k}\rangle _{j}^{\mathrm{ind}}$ are finite and uniquely defined. Our
main concern in the discussion below is that part of the energy-momentum
tensor. Integrating over the angular coordinates of $\mathbf{k}$ and
introducing the polar coordinates in the plane $(k,u)$, we integrate over
the related polar angle:%
\begin{eqnarray}
\langle \tau _{i}^{k}\rangle _{j}^{\mathrm{ind}} &=&\frac{(4\xi -1)\delta
_{i}^{k}\beta _{j}z^{D+1}}{2^{D-2}\pi ^{\frac{D-1}{2}}\Gamma (\frac{D-1}{2}%
)\alpha ^{D}}\mathbf{\,}\int_{0}^{\infty }d\gamma \,\gamma J_{\nu
}^{2}(\gamma z)\int_{0}^{\infty }dr\,r^{D-2}\,\frac{\beta _{j^{\prime
}}\lambda +1}{\beta _{j}\lambda -1}  \notag \\
&&\times \left. \frac{\lambda }{\left( \beta _{1}\lambda -1\right) \left(
\beta _{2}\lambda -1\right) e^{2a\lambda }-\left( \beta _{1}\lambda
+1\right) \left( \beta _{2}\lambda +1\right) }\right\vert _{\lambda =\sqrt{%
\gamma ^{2}+r^{2}}}.  \label{tauind1}
\end{eqnarray}%
Next we introduce polar coordinates in the plane $(\gamma ,r)$. The angular
integral is evaluated by using the result \cite{Prud2}%
\begin{equation}
\int_{0}^{1}dxx(1-x^{2})^{\frac{D-3}{2}}J_{\nu }^{2}(ux)=\frac{\Gamma (\frac{%
D-1}{2})}{2^{2\nu +1}}u^{2\nu }F_{\nu }(u),  \label{IntJ}
\end{equation}%
with the function%
\begin{equation}
F_{\nu }(u)=\frac{\,_{1}F_{2}(\nu +\frac{1}{2};\frac{D+1}{2}+\nu ,1+2\nu
;-u^{2})}{\Gamma (\frac{D+1}{2}+\nu )\Gamma (1+\nu )}.  \label{IntF}
\end{equation}%
Here, $_{1}F_{2}(a;b,c;x)$ is the hypergeometric function. This gives%
\begin{equation}
\langle \tau _{i}^{k}\rangle _{j}^{\mathrm{ind}}=\frac{(4\xi -1)\delta
_{i}^{k}\beta _{j}z^{D+2\nu +1}}{2^{D+2\nu -1}\pi ^{\frac{D-1}{2}}\alpha ^{D}%
}\mathbf{\,}\int_{0}^{\infty }d\lambda \,\frac{\beta _{j^{\prime }}\lambda +1%
}{\beta _{j}\lambda -1}\frac{\lambda ^{D+2\nu +1}F_{\nu }(\lambda z)}{\left(
\beta _{1}\lambda -1\right) \left( \beta _{2}\lambda -1\right) e^{2a\lambda
}-\left( \beta _{1}\lambda +1\right) \left( \beta _{2}\lambda +1\right) }.
\label{tauind3}
\end{equation}%
From here it follows that the induced SEMT on the brane $x^{1}=a_{j}$
vanishes for special cases of Dirichlet and Neumann boundary conditions on
that brane. Depending on the coefficients $\beta _{j}$ and on the separation
between the branes, the induced energy density $\varepsilon _{j}^{\mathrm{ind%
}}$ can be either positive or negative (see numerical examples below).
Introducing a new integration variable $u=\lambda z$, we see that the
product $\alpha ^{D}\langle \tau _{i}^{k}\rangle _{j}^{\mathrm{ind}}$
depends on the quantities $z$, $a_{j}$, $\beta _{j}$, having dimension of
length, in the form of two dimensionless ratios $a/z$, $\beta _{j}/z$. Those
ratios are the proper values of the quantities, measured by an observer with
fixed $z$, in units of the curvature radius $\alpha $. This feature is a
consequence of the AdS maximal symmetry.

\subsection{Minkowskian limit and a conformally coupled massless field}

To clarify the features of the SEMT on the branes we consider special cases
and asymptotic regions of the parameters. First we discuss the Minkowskian
limit corresponding to $\alpha \rightarrow \infty $ for fixed coordinate $y$%
. For the coordinate $z$, in the leading order, one has $z\approx \alpha $
and the line element (\ref{dsz}) tends to the Minkowskian interval $ds_{%
\mathrm{M}}^{2}=dt^{2}-\left( dx^{1}\right) ^{2}-d\mathbf{x}^{2}-dy^{2}$.
The geometry of the corresponding problem consists two parallel plates at $%
x^{1}=a_{1}$ and $x^{1}=a_{2}$ with the boundary condition $(1-(-1)^{j}\beta
_{j}\partial _{1})\varphi (x)=0$ at $x^{1}=a_{j}$ in the region $a_{1}\leq
x^{1}\leq a_{2}$. For large values of $\alpha $ and for a massive field the
parameter $\nu $ is large, $\nu \approx m\alpha $, and one needs the
asymptotic of the function $F_{\nu }(\lambda z)$ when both the argument and
the order are large. The respective analysis given in \cite{Beze15} shows
that the function $F_{\nu }(\nu \lambda /m)$ is exponentially suppressed for
$\nu \gg 1$ and $\lambda <m$. For $\lambda >m$ the leading behavior is
approximated by \cite{Beze15}%
\begin{equation}
F_{\nu }\left( \frac{\nu }{m}\lambda \right) \approx \frac{\left( \lambda
^{2}-m^{2}\right) ^{\frac{D}{2}-1}(2m/\nu )^{2\nu +1}}{2\sqrt{\pi }\Gamma (%
\frac{D}{2})\lambda ^{D+2\nu -1}}.  \label{Fnuas}
\end{equation}%
By using this asymptotic for the part of the integral in (\ref{tauind3})
over the region $m\leq \lambda <\infty $, one obtains the SEMT on the plate $%
x^{1}=a_{j}$ in Minkowski spacetime, $\langle \tau _{i}^{k}\rangle _{\mathrm{%
(M)}j}^{\mathrm{ind}}=\lim_{\alpha \rightarrow \infty }\langle \tau
_{i}^{k}\rangle _{j}^{\mathrm{ind}}$, given by%
\begin{equation}
\langle \tau _{i}^{k}\rangle _{\mathrm{(M)}j}^{\mathrm{ind}}=\frac{(4\xi
-1)\delta _{i}^{k}\beta _{j}}{2^{D-1}\pi ^{\frac{D}{2}}\Gamma (\frac{D}{2})}%
\mathbf{\,}\int_{m}^{\infty }d\lambda \,\frac{\beta _{j^{\prime }}\lambda +1%
}{\beta _{j}\lambda -1}\frac{\lambda ^{2}\left( \lambda ^{2}-m^{2}\right) ^{%
\frac{D}{2}-1}}{\left( \beta _{1}\lambda -1\right) \left( \beta _{2}\lambda
-1\right) e^{2a\lambda }-\left( \beta _{1}\lambda +1\right) \left( \beta
_{2}\lambda +1\right) }.  \label{SemtM0m}
\end{equation}%
This result for a massive field was obtained in \cite{Saha04cc} as a
limiting case of the problem with two branes in AdS spacetime parallel to
the AdS boundary. In the case of a massless field, the expression for $%
\langle \tau _{i}^{k}\rangle _{\mathrm{(M)}1}^{\mathrm{ind}}+\langle \tau
_{i}^{k}\rangle _{\mathrm{(M)}2}^{\mathrm{ind}}$, obtained from (\ref%
{SemtM0m}), coincides with the result derived in \cite{Rome02}. The VEV of
the SEMT for a single Robin boundary in background of (3+1)-dimensional
Minkowski spacetime has also been considered in \cite{Lebe96,Lebe01}.

In the case of a massless field with conformal coupling one has $\xi =\xi
_{D}=\frac{D-1}{4D}$ and $\nu =1/2$. By taking into account that $J_{1/2}(x)=%
\sqrt{\frac{\pi }{2x}}\sin x$, from (\ref{IntJ}) we get \cite{Beze15}%
\begin{equation}
F_{1/2}(u)=\frac{2}{\sqrt{\pi }u^{2}}\left[ \frac{1}{\Gamma \left( \frac{D}{2%
}\right) }-\frac{J_{\frac{D}{2}-1}(2u)}{u^{\frac{D}{2}-1}}\right] .
\label{Fcc}
\end{equation}%
Substituting this expression in (\ref{tauind3}) we get%
\begin{equation}
\varepsilon _{j}^{\mathrm{ind}}=\left( z/\alpha \right) ^{D}\varepsilon _{%
\mathrm{(M)}j}^{\mathrm{ind}},  \label{epscc}
\end{equation}%
with%
\begin{eqnarray}
\varepsilon _{\mathrm{(M)}j}^{\mathrm{ind}} &=&-\frac{2^{1-D}\beta _{j}}{%
D\pi ^{\frac{D}{2}}}\mathbf{\,}\int_{0}^{\infty }d\lambda \,\frac{\beta
_{j^{\prime }}\lambda +1}{\beta _{j}\lambda -1}\left[ \frac{1}{\Gamma \left(
\frac{D}{2}\right) }-\frac{J_{\frac{D}{2}-1}(2\lambda z)}{\left( \lambda
z\right) ^{\frac{D}{2}-1}}\right]   \notag \\
&&\times \frac{\lambda ^{D}}{\left( \beta _{1}\lambda -1\right) \left( \beta
_{2}\lambda -1\right) e^{2a\lambda }-\left( \beta _{1}\lambda +1\right)
\left( \beta _{2}\lambda +1\right) }.  \label{epsM}
\end{eqnarray}%
For a conformally coupled massless scalar field the problem we consider is
conformally related to the problem of two Robin plates at $x^{1}=a_{j}$, $%
j=1,2$, in Minkowski spacetime described by the interval $ds_{\mathrm{M}%
}^{2}=dt^{2}-\left( dx^{1}\right) ^{2}-d\mathbf{x}^{2}-dz^{2}$, intersected
by a Dirichlet plate located at $z=0$. The presence of the latter is related
to the boundary condition for scalar field modes imposed on the AdS boundary
$z=0$. The surface energy density (\ref{epsM}) is induced on the plate $%
x^{1}=a_{j}$ by the presence of the second plate $x^{1}=a_{j^{\prime }}$.
The part of $\varepsilon _{\mathrm{(M)}j}^{\mathrm{ind}}$ coming from the
first term in the square brackets is the respective quantity in the geometry
where the plate $z=0$ is absent (see (\ref{SemtM0m}) for $m=0$). The part
with the second term is a consequence of the presence of the plate $z=0$.
Note that $\varepsilon _{\mathrm{(M)}j}^{\mathrm{ind}}$ vanishes on that
plate: $\varepsilon _{\mathrm{(M)}j}^{\mathrm{ind}}|_{z=0}=0$. This is a
consequence of Dirichlet boundary condition at $z=0$.

\section{Asymptotics and numerical analysis}

\label{sec:Asymp}

In this section, the behavior of the VEV for SEMT in asymptotic regions of
the parameters is studied. We start with the asymptotics at small and large
separations between the branes. For a given $z$, the proper separation
between the branes is given by $a_{(p)}=\alpha a/z$. For small proper
separations compared to the curvature radius one has $a/z\ll 1$ and the
integral in (\ref{tauind3}) is dominated by the contribution of the region
with large values of the argument of the function $F_{\nu }(\lambda z)$. By
using the corresponding asymptotic \cite{Beze15}
\begin{equation}
F_{\nu }(u)\approx \frac{2^{2\nu }}{\sqrt{\pi }\Gamma \left( \frac{D}{2}%
\right) u^{2\nu +1}},\;u\gg 1,  \label{FnuLarge}
\end{equation}%
we can see that the relation
\begin{equation}
\langle \tau _{i}^{k}\rangle _{j}^{\mathrm{ind}}\approx \left( z/\alpha
\right) ^{D}\langle \tau _{i}^{k}\rangle _{\mathrm{(M)}j}^{\mathrm{ind}%
}|_{m=0},  \label{tausmalla}
\end{equation}%
takes place, where $\langle \tau _{i}^{k}\rangle _{\mathrm{(M)}j}^{\mathrm{%
ind}}|_{m=0}$ is given by (\ref{SemtM0m}) with $m=0$. In the limit under
consideration the main contribution to the SEMT comes from the zero-point
fluctuations with wavelengths smaller than the curvature radius and the
effect of the gravitational field is weak. The asymptotic (\ref{tausmalla})
is further simplified if the separation $a$ is smaller than the length
scales determined by the boundary conditions, $a/|\beta _{l}|\ll 1$, $l=1,2$%
. For Dirichlet boundary condition on the brane $x^{1}=a_{j^{\prime }}$, $%
\beta _{j^{\prime }}=0$, the condition $a/|\beta _{j}|\ll 1$ is assumed.
Under those conditions we have $\lambda |\beta _{l}|\gg 1$ ($\lambda |\beta
_{j}|\gg 1$ in the case $\beta _{j^{\prime }}=0$) for the region of $\lambda
$ that dominates in the integral on the right-hand side of (\ref{SemtM0m})
(with $m=0$). In the leading order one gets%
\begin{equation}
\langle \tau _{i}^{k}\rangle _{j}^{\mathrm{ind}}\approx \delta _{i}^{k}\frac{%
\left( z/\alpha \right) ^{D}(4\xi -1)}{2^{D}\pi ^{\frac{D+1}{2}}a^{D-1}}%
\zeta \left( D-1\right) \Gamma \left( \frac{D-1}{2}\right) \left\{
\begin{array}{ll}
1/\beta _{j^{\prime }}, & \beta _{j^{\prime }}\neq 0 \\
\left( 2^{2-D}-1\right) /\beta _{j}, & \beta _{j^{\prime }}=0%
\end{array}%
\right. ,  \label{tausmalla2}
\end{equation}%
with $\zeta \left( u\right) $ being the Riemann zeta function. Note that the
asymptotic (\ref{tausmalla}) also describes the behavior of the SEMT near
the AdS horizon. As it is seen from (\ref{tausmalla2}), in the special cases
of minimally ($\xi =0$) and conformally ($\xi =\xi _{D}$) coupled fields and
for small separations between the branes the energy density induced on the
brane $x^{1}=a_{j}$ by the second brane is positive for $\beta _{j^{\prime
}}<0$ and negative for $\beta _{j^{\prime }}>0$. For Dirichlet boundary
condition on the second brane ($\beta _{j^{\prime }}=0$) the sign of the
induced energy density coincides with the sign of the product $\left( 1-4\xi
\right) \beta _{j}$.

In the opposite limit of large proper separations compared with the
curvature radius, we have $a/z\gg 1$ and the main contribution to the
integral in (\ref{tauind3}) gives the region near the lower limit,
corresponding to $\lambda z\ll 1$. In the leading order, replacing the
function $F_{\nu }(\lambda z)$ by
\begin{equation}
F_{\nu }(0)=\frac{1}{\Gamma \left( \nu +1\right) \Gamma \left( \frac{D+1}{2}%
+\nu \right) },  \label{Fnu0}
\end{equation}%
one gets%
\begin{equation}
\langle \tau _{i}^{k}\rangle _{j}^{\mathrm{ind}}\approx \frac{8(4\xi
-1)\delta _{i}^{k}\left( z/2\right) ^{D+2\nu +2}\beta _{j}/z}{\pi ^{\frac{D-1%
}{2}}\Gamma \left( \nu +1\right) \Gamma \left( \frac{D+1}{2}+\nu \right)
\alpha ^{D}}\mathbf{\,}\int_{0}^{\infty }d\lambda \,\frac{\lambda \beta
_{j^{\prime }}+1}{\lambda \beta _{j}-1}\frac{\lambda ^{D+2\nu +1}}{\left(
\lambda \beta _{1}-1\right) \left( \lambda \beta _{2}-1\right) e^{2\lambda
a}-\left( \lambda \beta _{1}+1\right) \left( \lambda \beta _{2}+1\right) }.
\label{taularge}
\end{equation}%
This expression is further simplified for separations larger than the length
scales in Robin boundary conditions. Assuming $a\gg |\beta _{l}|$, $l=1,2$,
we see that $\lambda |\beta _{l}|\ll 1$ for the region giving the dominant
contribution to the integral in (\ref{taularge}). For the case of Neumann
boundary condition on the brane $x^{1}=a_{j^{\prime }}$, corresponding to
the limit $|\beta _{j}|\rightarrow \infty $, for separations $a\gg |\beta
_{j}|$ one has $\lambda |\beta _{j}|\ll 1$ in the region with dominant
contribution to the integral. For the leading order term in the VEV of the
SEMT and for non-Neumann boundary conditions on the second brane we find
\begin{equation}
\langle \tau _{i}^{k}\rangle _{j}^{\mathrm{ind}}\approx \delta _{i}^{k}\frac{%
(1-4\xi )\zeta \left( D+2\nu +2\right) \beta _{j}/a}{\pi ^{\frac{D}{2}%
}\Gamma \left( \nu +1\right) \alpha ^{D}\left( 2a/z\right) ^{D+2\nu +1}}%
\mathbf{\,}\left( D+2\nu +1\right) \Gamma \left( \frac{D}{2}+\nu +1\right) .
\label{taularge2}
\end{equation}%
For Neumann boundary condition on the second brane, an additional factor $%
(2^{-D-2\nu -1}-1)$ should be added in the right-hand side of (\ref%
{taularge2}). We see that at large distances between the branes the decay of
the SEMT, as a function of the proper separation, is a power law for both
massive and massless fields. This feature for massive fields is in contrast
with the corresponding behavior for parallel plates in the Minkowski bulk,
where the suppression is exponential, by the factor $e^{-2ma}$. We note that
the formula (\ref{taularge}) also gives the asymptotic of the SEMT near the
AdS boundary. As seen, for fixed $\beta _{j}$, the SEMT tends to zero on the
AdS boundary like $z^{D+2\nu +1}$. The asymptotic estimate (\ref{taularge2})
shows that for $\beta _{j}<0$ and for non-Neumann boundary conditions on the
second brane ($1/\beta _{j^{\prime }}\neq 0$), at large separations between
the branes the induced energy density $\varepsilon _{j}^{\mathrm{ind}}$ is
negative for minimally and conformally coupled fields.

\begin{figure}[tbph]
\begin{center}
\begin{tabular}{cc}
\epsfig{figure=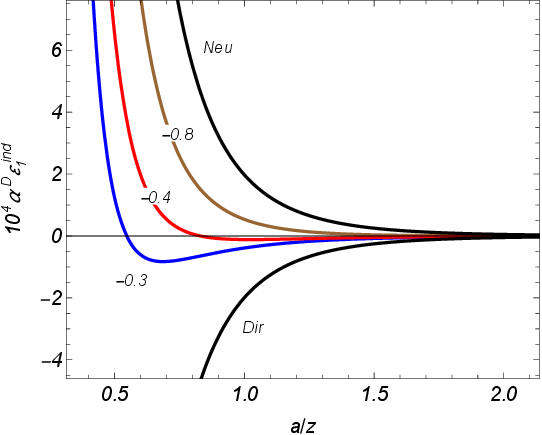,width=7.5cm,height=6cm} & \quad %
\epsfig{figure=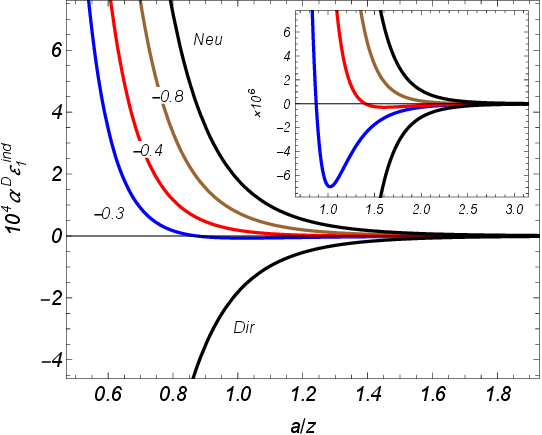,width=7.5cm,height=6cm}%
\end{tabular}%
\end{center}
\caption{The induced surface energy density on the brane at $x^{1}=a_{1}$,
in units of $\protect\alpha ^{-D}$, versus the proper separation between the
branes for $D=4$, $m\protect\alpha =0.5$, and $\protect\beta _{1}/z=0.5$.
The graphs are presented for different values of the ratio $\protect\beta %
_{2}/z$ (the numbers near the curves) and for Dirichlet and Neumann boundary
conditions on the second brane ($\protect\beta _{2}/z=0$ and $\protect\beta %
_{2}/z=\infty $, respectively). }
\label{figdist}
\end{figure}

Figure \ref{figdist} presents the VEV of the energy density, induced on the
brane at $x^{1}=a_{1}$ by the brane at $x^{1}=a_{2}$, as a function of the
proper separation between the branes $a/z$. The graphs are plotted for a
scalar field in (4+1)-dimensional AdS spacetime ($D=4$), for Robin boundary
condition with $\beta _{1}/z=-0.5$ and with the mass corresponding to $%
m\alpha =0.5$. The dependence on the proper separation is displayed for
different values of the ratio $\beta _{2}/z$ (the numbers near the curves)
and for Dirichlet and Neumann boundary conditions on the second brane. The
left and right panels correspond to conformally and minimally coupled
fields, respectively. In accordance with the asymptotic analysis given
above, for minimally and conformally coupled fields and at small separations
between the branes, the energy density, induced by the second brane, is
positive (negative) for non-Dirichlet (Dirichlet) boundary condition on the
second brane. At large separations the energy density is negative for
non-Neumann boundary conditions on the second brane and is positive for
Neumann boundary condition. The inset on the right panel of Figure \ref{figdist}
is presented to clearly emphasize the change in the sign of the surface energy density
as a function of the separation between the branes.

\begin{figure}[tbph]
\begin{center}
\begin{tabular}{cc}
\epsfig{figure=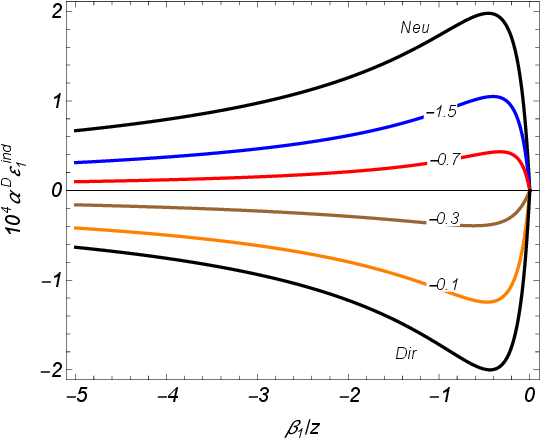,width=7.5cm,height=6cm} & \quad %
\epsfig{figure=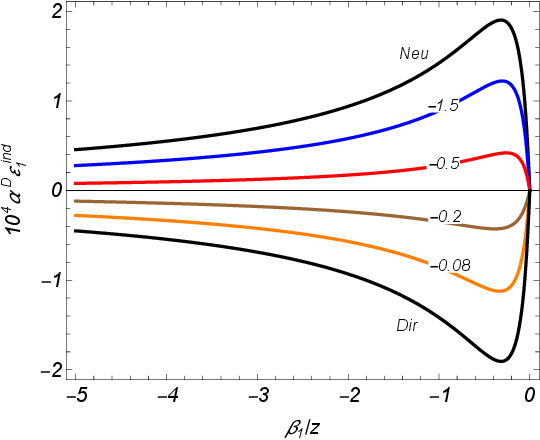,width=7.5cm,height=6cm}%
\end{tabular}%
\end{center}
\caption{The induced surface energy density on the brane at $x^{1}=a_{1}$
versus the Robin coefficient $\protect\beta _{1}/z$ for different values of $%
\protect\beta _{2}/z$ (the numbers near the curves, $\protect\beta _{2}/z=0$
and $\protect\beta _{2}/z=-\infty $ for Dirichlet and Neumann conditions).
The graphs are plotted for conformally (left panel) and minimally (right
panel) coupled fields and for $D=4$, $m\protect\alpha =0.5$, and $a/z=1$. }
\label{figbeta}
\end{figure}

In Figure \ref{figbeta}, for conformally (left panel) and minimally (right
panel) coupled scalar fields in $D=4$ spatial dimensions, we have plotted
the dependence of the energy density $\varepsilon _{1}^{\mathrm{ind}}$ on
the Robin coefficient $\beta _{1}/z$ for different values of the Robin
coefficient $\beta _{2}/z$ on the second brane (the numbers near the curves)
and for Dirichlet and Neumann boundary conditions. The graphs are plotted
for $m\alpha =0.5$ and $a/z=1$.

\begin{figure}[tbph]
\begin{center}
\begin{tabular}{cc}
\epsfig{figure=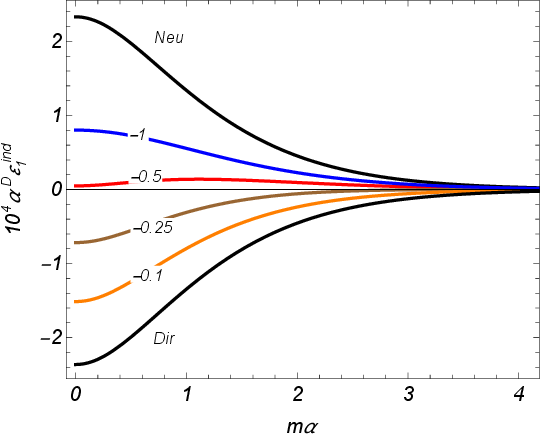,width=7.5cm,height=6cm} & \quad %
\epsfig{figure=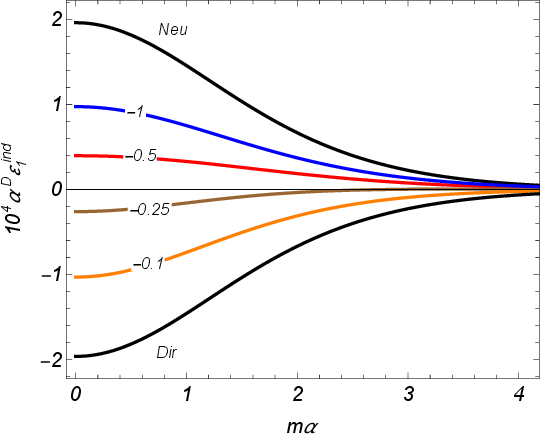,width=7.5cm,height=6cm}%
\end{tabular}%
\end{center}
\caption{The dependence of the surface energy density on the first brane,
induced by the second brane, versus the field mass for conformally and
minimally coupled fields (left and right panels, respectively). The graphs
are plotted for $D=4$, $a/z=1$, $\protect\beta _{1}/z=-0.5$ and for separate
values of $\protect\beta _{2}/z$ (the numbers near the curves). The graphs
for Dirichlet and Neumann boundary conditions on the second brane are
presented as well.}
\label{figmass}
\end{figure}

The dependence of the surface energy density on the mass of the field (in
units of $1/\alpha $) is displayed in Figure \ref{figmass} for conformally
(left panel) and minimally (right panel) coupled scalar field in spatial
dimensions $D=4$. The graphs are plotted for $a/z=1$, $\beta _{1}/z=-0.5$,
and for different values of the ratio $\beta _{2}/z$ (the numbers near the
curves). The graphs corresponding to Robin boundary conditions, $-\infty
<\beta _{2}/z<0$, are located between the graphs corresponding to Neumann
and Dirichlet boundary conditions on the second brane ($\beta _{2}/z=-\infty
$ and $\beta _{2}/z=0$, respectively). As seen, the induced energy density,
in general, is not a monotonic function of the field mass. In addition, for
fixed values of the other parameters it may change the sign as a function of
the mass. In particular, that is the case for a minimally coupled field with
the boundary conditions corresponding to $\beta _{1}/z=-0.5$ and $\beta
_{2}/z=-0.25$ (see the right panel in Figure \ref{figmass}).

\section{Conclusion}

\label{sec:Conc}

For a scalar field with general curvature coupling, we have studied the VEV
of the SEMT induced on branes in AdS spacetime orthogonal to its boundary.
On the branes the field operator is constrained by the boundary conditions (%
\ref{Rbc}) or, equivalently, by (\ref{Rbc2}). To ensure the stability of the
vacuum state, the values of the parameters in Robin boundary conditions are
restricted by (\ref{StabCond}). For the geometry of the branes under
consideration the extrinsic curvature tensor is zero and the general formula
for the SEMT is simplified to (\ref{tauikvev2}). From the viewpoint of
observers living on the branes this SEMT presents a gravitational source
with the equation of state for a cosmological constant. In order to evaluate
the corresponding VEV the Hadamard function is used that is obtained from
the positive frequency Wightman function from \cite{Bell22}. In the region
between the branes the Hadamard function is decomposed into single brane and
the second brane induced contributions. That allows to separate from the
total VEV of the SEMT the part generated by the second brane. The surface
divergences are contained in the self-energy contributions on the branes and
the renormalization is required for those parts only. In order to extract
the finite parts in the respective VEVs, in Appendix \ref{sec:App} we have
employed the regularization procedure based on the generalized zeta function
approach. The divergences appearing in the form of simple poles are absorbed
by the renormalization of the respective parameters in the "classical"
action localized on the branes. The finite part of the SEMT separated in
this way contains renormalization ambiguities and additional conditions are
required to obtain a unique result. Here the situation is completely
parallel to the one for the self-energy in the Casimir effect in the
geometry of a single boundary (see, for example, the respective discussion
in \cite{Bord09}).

The part of the SEMT induced on the brane by the presence of the second
brane is finite and uniquely defined. The induced SEMT on the brane $%
x^{1}=a_{j}$ is given by the expression (\ref{tauind3}). It vanishes for
special cases of Dirichlet and Neumann boundary conditions on that brane. As
a consequence of the maximal symmetry of AdS spacetime, for general case of
Robin boundary conditions the dimensionless quantity $\alpha ^{D}\langle
\tau _{i}^{k}\rangle _{j}^{\mathrm{ind}}$ is completely determined by
dimensionless ratios $a/z$ and $\beta _{j}/z$, $j=1,2$. The first one is the
proper separation between the branes, measured by an observer with fixed $z$
in units of the curvature radius $\alpha $. The VEV\ of the SEMT for Robin
parallel plates in Minkowski bulk is obtained from (\ref{tauind3}) in the
limit $\alpha \rightarrow \infty $ and is expressed as (\ref{SemtM0m}). The
latter includes special cases previously discussed in the literature and
coincides with the result obtained in \cite{Saha04cc} as a limit $\alpha
\rightarrow \infty $ of the SEMT in the geometry of branes parallel to the
AdS boundary. For a conformally coupled massless field the problem in the
AdS bulk is conformally related to the problem in Minkowski spacetime
consisting two parallel Robin plates perpendicularly intersected by a
Dirichlet plate, the latter being the image of the AdS boundary. The VEV in
the Minkowski counterpart is given by the formula (\ref{epsM}), where the
contribution of the Dirichlet plate comes from the term in the square
brackets with the Bessel function.

At small separations between the branes, compared to the curvature radius
and length scales determined by the Robin coefficients, the influence of the
gravitational field on the SEMT is small and the leading term in the
respective expansion is expressed by (\ref{tausmalla2}). In this limit and
for non-Dirichlet (Dirichlet) boundary conditions on the brane $%
x^{1}=a_{j^{\prime }}$ the sign of the surface energy density induced on the
brane $x^{1}=a_{j}$ coincides with the sign of the product $(4\xi -1)\beta
_{j^{\prime }}$ ($(1-4\xi )\beta _{j}$). The effects of the gravitational
field are essential at proper separations between the branes of the order or
larger than the curvature scale of the background geometry. Additionally
assuming that the separation is large than the length scales fixed by
boundary conditions, the leading behavior of the induced SEMT is described
by (\ref{taularge2}) for non-Neumann boundary conditions on the second
brane. The sign of the energy density coincides with the sign of $\left(
1-4\xi \right) \beta _{j}$. For Neumann condition on the second brane, an
additional factor $(2^{-D-2\nu -1}-1)$ should be added in the right-hand
side and the energy density at large distances has an opposite sign. An
important feature of the large distance behavior of the SEMT is the power
law decay as a function of the proper separation. For parallel plates in
Minkowski spacetime the respective decay for massive fields is exponential.
The induced surface energy density vanishes on the AdS boundary like $%
z^{D+2\nu +1}$ and behaves as $\left( z/\alpha \right) ^{D}$ near the AdS
horizon.

The investigations of the brane induced effects on the properties of the scalar vacuum
in AdS spacetime have discussed the branes parallel or perpendicular to the AdS boundary.
An interesting generalization, that includes these special cases, would be the geometry of branes
crossing the AdS boundary at an arbitrary angle. In this case, the dependence of the scalar mode functions
on the coordinates parallel and perpendicular to the AdS boundary are not separable and the problem
is more complicated. It is expected that for a general crossing angle, in addition the normal and shear Casimir forces,
a rotational momentum will appear generated by the vacuum fluctuations.

The study of the boundary induced effects on the fermionic and electromagnetic vacua for branes perpendicular to the AdS boundary is another direction
for further research. The dependence of the mode functions on the coordinate $z$ is expressed in terms of the
functions $J_{m \alpha \pm 1/2}(\gamma z)$ for the fermionic field (with $m$ being the mass of the field)
and in terms of the function $J_{D/2-1}(\gamma z)$ for the vector potential of the electromagnetic field.
Similar to the case of a scalar field, we expect that the equation determining the eigenvalues of the quantum number
corresponding to the direction normal to the branes will be the same as that in the Minkowski bulk with the same boundary conditions
on planar boundaries. The summation over those eigenvalues in the respective mode sum for the VEV of the energy-momentum tensor can
be done by using the generalized Abel-Plana formula. That allows to extract explicitly the brane induced contribution.
Note that the previous investigations of the vacuum energy-momentum tensor for fermionic and electromagnetic fields have considered
branes parallel to the AdS boundary (see references \cite{Shao10}-\cite{Saha20}). The bag boundary condition has been
imposed for the fermionic field and for the electromagnetic field the perfect conductor and confining boundary conditions
have been discussed.

\section*{Acknowledgments}

The work was supported by the grant No. 21AG-1C047 of the Higher Education
and Science Committee of the Ministry of Education, Science, Culture and
Sport RA.

\appendix

\section{Surface densities for a single brane}

\label{sec:App}

We have seen that the VEV of the SEMT for a single brane at $x^{1}=a_{j}$ is
presented in the form (\ref{tau0}). The corresponding expression is
divergent and we can regularize it by using the generalized zeta function
approach (for a general introduction and applications in the theory of the
Casimir effect see, e.g., \cite{Eliz94,Kirs02,Eliz12}). Let us consider the
function%
\begin{equation}
F(s,z)=\frac{\mu ^{s-1}\beta _{j}z^{D+1}}{(2\pi )^{D-1}}\int_{0}^{\infty
}d\gamma \,\gamma J_{\nu }^{2}(\gamma z)\int_{0}^{\infty }d\lambda \,\lambda
^{2}\int d\mathbf{k\,}\frac{\left( \lambda ^{2}+\gamma ^{2}+k^{2}\right) ^{-%
\frac{s}{2}}}{1+\lambda ^{2}\beta _{j}^{2}},  \label{Fsz}
\end{equation}%
with, in general, complex argument $s$. As it will be seen below, the
expression in the right-hand is finite for $\mathrm{Re}\,s>D$. The scale
parameter $\mu $, having dimension of inverse length, is introduced to keep
the function $F(s,z)$ dimensionless. Following the principal part
prescription, considered previously in the literature for the total Casimir
energy in ultrastatic manifolds with boundaries (see, \cite%
{Eliz94,Kirs02,Blau88}), the SEMT in the geometry of a single brane is
obtained as%
\begin{equation}
\langle \tau _{i}^{k}\rangle _{j}^{(0)}=\delta _{i}^{k}\frac{4\xi -1}{\alpha
^{D}}\mathrm{PP}\left[ F(s,z)\right] _{s=1},  \label{analytic}
\end{equation}%
where $\mathrm{PP}\left[ F(s,z)\right] _{s=1}$ corresponds to the finite
part of the Laurent expansion of the function $F(s,z)$ near $s=1$. The
evaluation of that part is reduced to the extraction of the pole term.

The integral over $\mathbf{k}$ in (\ref{Fsz}) is expressed in terms of the
gamma function and we get%
\begin{equation}
F(s,z)=\frac{\mu ^{s-1}\beta _{j}z^{D+1}}{2^{D-1}\pi ^{D/2}}\frac{\Gamma (1-%
\frac{D-s}{2})}{\Gamma (\frac{s}{2})}\int_{0}^{\infty }d\gamma \,\gamma
J_{\nu }^{2}(\gamma z)\int_{0}^{\infty }d\lambda \,\lambda ^{2}\frac{\left(
\lambda ^{2}+\gamma ^{2}\right) ^{\frac{D-s}{2}-1}}{1+\lambda ^{2}\beta
_{j}^{2}}.  \label{Fsz1}
\end{equation}%
For the further transformation of the expression in the right-hand side of (%
\ref{Fsz1}) we use the integral representation%
\begin{equation}
\left( \lambda ^{2}+\gamma ^{2}\right) ^{\frac{D-s}{2}-1}=\frac{1}{\Gamma
\left( 1-\frac{D-s}{2}\right) }\int_{0}^{\infty }dx\,x^{\frac{s-D}{2}%
}e^{-\left( \lambda ^{2}+\gamma ^{2}\right) x}.  \label{IntRep}
\end{equation}%
With this representation, the integral over $\gamma $ is evaluated by the
formula \cite{Prud2}:
\begin{equation}
\int_{0}^{\infty }d\gamma \,\gamma J_{\nu }^{2}(\gamma z)e^{-\gamma ^{2}x}=%
\frac{1}{2x}\exp \left( -\frac{z^{2}}{2x}\right) I_{\nu }\left( \frac{z^{2}}{%
2x}\right) ,  \label{IntJ2}
\end{equation}%
with $I_{\nu }\left( u\right) $ being the modified Bessel function. Passing
to a new integration variable $u=z^{2}/(2x)$, one finds%
\begin{equation}
F(s,z)=\frac{\mu ^{s-1}\beta _{j}z^{s+1}}{2^{\frac{D+s}{2}}\pi ^{D/2}\Gamma (%
\frac{s}{2})}\int_{0}^{\infty }du\,u^{\frac{D-s}{2}-1}e^{-u}I_{\nu }\left(
u\right) \int_{0}^{\infty }d\lambda \,\frac{\lambda ^{2}e^{-\lambda ^{2}%
\frac{z^{2}}{2y}}}{1+\lambda ^{2}\beta _{j}^{2}}.  \label{Fsz2}
\end{equation}%
The $\lambda $-integral is evaluated in terms of the complementary
incomplete gamma function $\Gamma (-1/2,x)$. As a result, the function $%
F(s,z)$ is presented as
\begin{equation}
F(s,z)=\frac{\left( \mu z\right) ^{s-1}\beta _{j}z^{2}}{2^{\frac{D+s}{2}%
+2}\pi ^{\frac{D-1}{2}}\Gamma (\frac{s}{2})|\beta _{j}|^{3}}\int_{0}^{\infty
}du\,u^{\frac{D-s}{2}-1}S\left( 2\beta _{j}^{2}/z^{2},u\right) ,
\label{Fsz3}
\end{equation}%
where we have introduced the function%
\begin{equation}
S(b,u)=e^{-u}I_{\nu }\left( u\right) e^{\frac{1}{bu}}\Gamma \left( -\frac{1}{%
2},\frac{1}{bu}\right) .  \label{Sbu}
\end{equation}%
In the limit $u\rightarrow \infty $ the function (\ref{Sbu}) tends to
limiting value $\sqrt{2b/\pi }$ and $\lim_{u\rightarrow 0}S(b,u)=0$. This
shows that the representation (\ref{Fsz3}) is valid in the region $\mathrm{Re%
}\,s>D$ of the complex plane $s$.

The divergence of the integral in (\ref{Sbu}) at $s=1$ comes from the
divergence in the upper limit of the integral. By using the expansions of
the functions $e^{-u}I_{\nu }\left( u\right) $ and $e^{\frac{1}{bu}}\Gamma
\left( -\frac{1}{2},\frac{1}{bu}\right) $ (see, for example, \cite{Abra72})
for large values of $u$, the following expansion is obtained:
\begin{equation}
S(b,u)=\sqrt{\frac{2b}{\pi }}\sum_{n=0}^{\infty }\left[ \frac{A_{n}(b)}{u^{n}%
}-\sqrt{\pi }\frac{B_{n}(b)}{u^{n+\frac{1}{2}}}\right] .  \label{Sbuexp}
\end{equation}%
For the coefficients one has%
\begin{eqnarray}
A_{0} &=&1,\;A_{1}=\frac{2}{b}-\frac{1}{2}\left( \nu ^{2}-\frac{1}{4}\right)
,  \notag \\
A_{2} &=&\frac{4}{3b^{2}}+\left( \nu ^{2}-\frac{1}{4}\right) \left[ \frac{1}{%
8}\left( \nu ^{2}-\frac{9}{4}\right) -\frac{1}{b}\right] ,  \label{A012}
\end{eqnarray}%
and
\begin{eqnarray}
B_{0} &=&\frac{1}{\sqrt{b}},\;B_{1}=\frac{1}{b^{\frac{3}{2}}}-\frac{1}{2%
\sqrt{b}}\left( \nu ^{2}-\frac{1}{4}\right) ,  \notag \\
B_{2} &=&\frac{1}{2\sqrt{b}}\left[ \frac{1}{b^{2}}+\left( \nu ^{2}-\frac{1}{4%
}\right) \left( \frac{1}{4}\left( \nu ^{2}-\frac{9}{4}\right) -\frac{1}{b}%
\right) \right] .  \label{B012}
\end{eqnarray}

In order to separate the pole term in (\ref{Fsz3}) we rewrite the function $%
F(s,z)$ in the form
\begin{eqnarray}
F(s,z) &=&\frac{\left( \mu z\right) ^{s-1}\beta _{j}z^{2}}{2^{\frac{D+s}{2}%
+2}\pi ^{\frac{D-1}{2}}\Gamma (\frac{s}{2})|\beta _{j}|^{3}}\left\{
\int_{0}^{1}du\,u^{\frac{D-s}{2}-1}S\left( b_{j},u\right) \right.  \notag \\
&&\left. +\int_{1}^{\infty }du\,u^{\frac{D-s}{2}-1}\left[ S\left(
b_{j},u\right) -S_{N}\left( b_{j},u\right) \right] +\int_{1}^{\infty }du\,u^{%
\frac{D-s}{2}-1}S_{N}\left( b_{j},u\right) \right\} ,  \label{Fsz4}
\end{eqnarray}%
where $b_{j}=2\beta _{j}^{2}/z^{2}$ and%
\begin{equation}
S_{N}(b,u)=\sqrt{\frac{2b}{\pi }}\sum_{n=0}^{N}\left[ \frac{A_{n}(b)}{u^{n}}-%
\sqrt{\pi }\frac{B_{n}(b)}{u^{n+\frac{1}{2}}}\right] .  \label{SN}
\end{equation}%
For $N>(D-3)/2$ the first two integrals in the figure braces (\ref{Fsz3})
are convergent for $s=1$. By using (\ref{SN}) in the part coming from the
last integral in (\ref{Fsz4}), the corresponding contribution to the
function $F(s,z)$ is presented as%
\begin{equation}
\bar{F}(s,z)=-\frac{\left( \mu z/\sqrt{2}\right) ^{s-1}z}{2^{\frac{D+1}{2}%
}\pi ^{\frac{D}{2}}\Gamma (\frac{s}{2})\beta _{j}}\sum_{n=0}^{N}\left[ \frac{%
A_{n}(b_{j})}{s+2n-D}-\frac{\sqrt{\pi }B_{n}(b_{j})}{s+1+2n-D}\right] .
\label{Ftsz}
\end{equation}%
The function $\bar{F}(s,z)$ has a simple pole at $s=1$. The pole comes from
the term with $n=(D-1)/2$ for odd $D$ and from the term with $n=D/2-1$ for
even $D$.

Expanding the function (\ref{Ftsz}) near the physical point $s=1$, the
function $F(s,z)$ is decomposed as%
\begin{equation}
F(s,z)=\frac{F_{\mathrm{(p)}}(s,z)}{s-1}+F_{\mathrm{(f)}}(z)+\cdots ,
\label{Fdec}
\end{equation}%
where the ellipsis stand for the part vanishing in the limit $s\rightarrow 1$%
. Here, the coefficient in the pole term and the finite part are given by
the expressions%
\begin{equation}
F_{\mathrm{(p)}}(s,z)=-\frac{zC_{D}(b_{j})}{\left( 2\pi \right) ^{\frac{D+1}{%
2}}\beta _{j}},  \label{Fp}
\end{equation}%
and%
\begin{eqnarray}
F_{\mathrm{(f)}}(z) &=&\frac{\beta _{j}z^{2}}{2^{\frac{D+1}{2}+2}\pi ^{\frac{%
D}{2}}|\beta _{j}|^{3}}\left\{ \int_{0}^{1}du\,u^{\frac{D-3}{2}}S\left(
b_{j},u\right) +\int_{1}^{\infty }du\,u^{\frac{D-3}{2}}\left[ S\left(
b_{j},u\right) -S_{N}\left( b_{j},u\right) \right] \right\}   \notag \\
&+&\frac{z}{\left( 2\pi \right) ^{\frac{D+1}{2}}\beta _{j}}\left\{
C_{D}(b_{j})\left[ \ln \left( \frac{\mu z}{\sqrt{2}}\right) +\frac{1}{2}\psi
(1/2)\right] -\sum_{n=0}^{N\prime }\left[ \frac{A_{n}(b_{j})}{1+2n-D}-\frac{%
\sqrt{\pi }B_{n}(b_{j})}{2+2n-D}\right] \right\} ,  \label{Ff}
\end{eqnarray}%
where the prime on the summation sign means that the term $n=\frac{D-1}{2}$
for odd $D$ and the term $n=\frac{D}{2}-1$ for even $D$ should be omitted.
In (\ref{Ff}), $\psi (x)$ is the digamma function with $\psi (1/2)\approx
-1.964$ and%
\begin{equation}
C_{D}(b)=\left\{
\begin{array}{ll}
A_{\frac{D-1}{2}}(b), & \text{for odd }D \\
-\sqrt{\pi }B_{\frac{D}{2}-1}(b) & \text{for even }D%
\end{array}%
\right. .  \label{CDb}
\end{equation}%
In the principal part prescription, the physical value extracted from the
divergent expectation value of the SEMT $\langle \tau _{i}^{k}\rangle
_{j}^{(0)}$ is identified with
\begin{equation}
\langle \tau _{i}^{k}\rangle _{j}^{(0)}=\delta _{i}^{k}\frac{4\xi -1}{\alpha
^{D}}F_{\mathrm{(f)}}(z).  \label{tau0fin}
\end{equation}%
Note that this result contains a scale ambiguity. Under scale change it
transforms as%
\begin{equation}
\langle \tau _{i}^{k}\rangle _{j}^{(0)}(\mu ^{\prime })=\langle \tau
_{i}^{k}\rangle _{j}^{(0)}(\mu )+\delta _{i}^{k}\left( 4\xi -1\right) \frac{%
\ln (\mu ^{\prime }/\mu )C_{D}(b_{j})z}{\left( 2\pi \right) ^{\frac{D+1}{2}%
}\alpha ^{D}\beta _{j}}.  \label{ScaleTrans}
\end{equation}%
The logarithmic dependence on the scale $\mu $ is a characteristic feature
of the regularization procedure.

\end{document}